\documentstyle[prl,aps,floats,epsf]{revtex}

\def\sgn{\mathop{\rm sgn}}
\def\heviside{\mathop{\theta}}

\begin{document} 
\draft
\twocolumn[\hsize\textwidth\columnwidth\hsize\csname
@twocolumnfalse\endcsname

\preprint{HEP/123-qed}

\title{Disorder Effects in the Bipolaron System Ti$_{4}$O$_{7}$ 
Studied by Photoemission Spectroscopy}

\author{K. Kobayashi\cite{presentaddress}, T. Susaki, A. Fujimori$^{a}$, T. Tonogai$^{b}$, and H. 
Takagi$^{c}$} 

\address{Department of Physics, University of Tokyo, Bunkyo-ku, Tokyo, 
113-0033, Japan}

\address{$^{a}$Department of Physics and Department of Complexity 
Science and Engineering, \\ University of Tokyo, Bunkyo-ku, Tokyo 
113-0033, Japan}

\address{$^{b}$Department of Applied Physics, University of Tokyo, 
Bunkyo-ku, Tokyo 113-0033, Japan}

\address{$^{c}$Department of Applied Chemistry and Department of 
Advanced Materials Science, \\ University of Tokyo, Bunkyo-ku, Tokyo 
113-0033, Japan}

\date{August 30, 1999}
\maketitle

\begin{abstract}
We have performed a photoemission study of Ti$_{4}$O$_{7}$ around its 
two transition temperatures so as to cover the metallic, 
high-temperature insulating (bipolaron-liquid), and low-temperature 
insulating (bipolaron-crystal) phases.  While the spectra of the 
low-temperature insulating phase show a finite gap at the Fermi level, 
the spectra of the high-temperature insulating phase are gapless, 
which is interpreted as a soft Coulomb gap due to dynamical disorder.  
We suggest that the spectra of the high-temperature disordered phase 
of Fe$_{3}$O$_{4}$, which exhibits a charge order-disorder transition 
(Verwey transition), can be interpreted in terms of a Coulomb gap.
\end{abstract}


\vskip1pc]
\narrowtext

Since Mott~\cite{MottMIT} proposed the idea of variable-range hopping 
and minimum metallic conductivity for disordered systems and 
Anderson~\cite{AndersonPRL1975} raised the concept of localization due 
to disorder, physical properties of disordered solids have been 
extensively studied.  Influence of Coulomb interaction on the 
electronic density of states (DOS) near the Fermi level ($E_{F}$) of 
disordered systems is one of the most fundamental issues to be 
clarified.  Efros and Shklovskii~\cite{EfrosJPC1975} proposed that in 
disordered insulators long-range Coulomb interaction opens a soft 
Coulomb gap (SCG), whose DOS is proportional to $(E-E_{F})^{2}$.  So 
far, there have been tunneling spectroscopic confirmations of the SCG 
in some disordered systems such as doped 
semiconductors~\cite{MasseyPRL1996}.  Davies and 
Franz~\cite{DaviesPRL1986} pointed out the possibility of an SCG 
opening in the photoemission spectra of 
Na$_{x}$Ta$_{y}$W$_{1-y}$O$_{3}$~\cite{HollingerPRB1985}, but the 
experiments did not have sufficient energy resolution to critically 
address this problem.  Another important issue is how short-range 
order (SRO) affects DOS near $E_{F}$, that is, how the electronic 
structure evolves when a charge ordering develops from disorder to SRO 
to long-range order.

\begin{figure} [ht] \center \epsfxsize=83mm \epsfbox{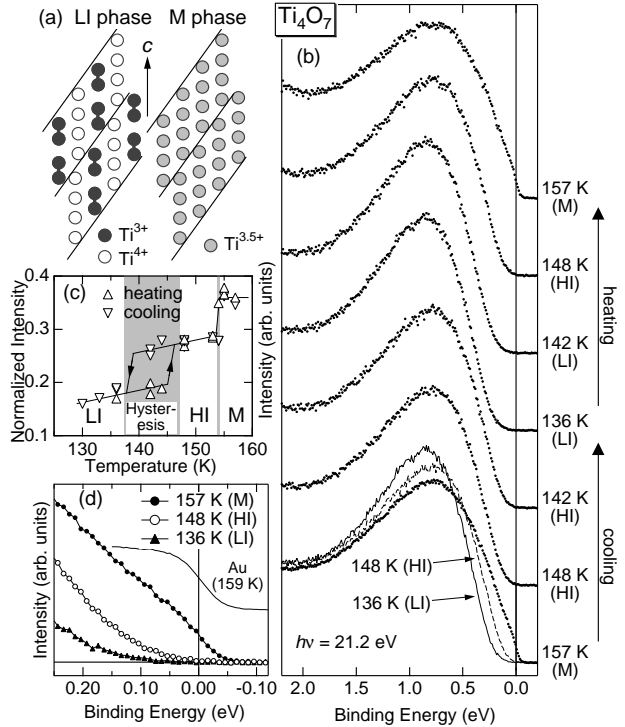}
\vspace{2pt} \caption{(a) Schematic crystal structures of 
Ti$_{4}$O$_{7}$ which show only the chains parallel to the 
pseudo-rutile \textit{c}-axis truncated by shear planes indicated by 
thick lines~\protect\cite{MarezioPRL1972}.  In the LI phase, the 
bipolarons are ordered.  In the M phase, all the Ti sites have a 
uniform valence of 3.5+.  (b) PES spectra of Ti$_{4}$O$_{7}$ taken by 
varying the temperature from the bottom to the top.  (c) Normalized 
PES intensity near $E_{F}$ (see text).  The solid line is guide to the 
eye.  (d) The M-, HI-, and LI-phase spectra around $E_{F}$ plotted on 
an expanded scale.}
\label{Ti4O7narrow}
\end{figure}

Ti$_{4}$O$_{7}$ is a suitable system to study the above problems: It 
undergoes successive phase transitions with decreasing temperature 
from a metal to a charge-ordered insulator via an insulator with 
SRO~\cite{SchlenkerPDM1985}.  It is a system with nominally 0.5 
3\textit{d} electron per Ti, allowing two possible valence states of 
Ti$^{3+}$ ($3d^{1}$) and Ti$^{4+}$ ($3d^{0}$), and attracted 
particular attention in 1970's as a system where bipolarons, or 
singlet pairs of two polarons, are formed in real 
space~\cite{SchlenkerPDM1985,MarezioPRL1972,SchlenkerPRL1974}.  Above 
$T_{MI} = 154$ K, the system is in the metallic (M) phase and the Ti 
valence is believed to be uniform 3.5+ as shown in 
Fig.~\ref{Ti4O7narrow} (a).  That is, the electrical resistivity 
$\rho(T)$ is metallic with Pauli-paramagnetic $\chi(T)$.  With 
decreasing temperature, singlet Ti$^{3+}$-Ti$^{3+}$ pairs, namely 
bipolarons, are formed, resulting in the metal-to-insulator transition 
at $T_{MI}$ with a steep increase in $\rho(T)$ by three orders of 
magnitude.  At the same time, $\chi(T)$ almost vanishes, reflecting 
the formation of the singlet bipolarons.  We refer to this phase as 
the high-temperature insulating (HI) phase.  Because the bipolarons 
are dynamically disordered in this phase as has been established by 
EPR studies~\cite{LakkisPRB1976}, the HI phase may be called a 
bipolaron liquid.  Subsequently, bipolarons become ordered below 
$T_{II} \sim 140$ K as shown in Fig.~\ref{Ti4O7narrow} (a), with a 
further rise in $\rho (T)$ by three orders of magnitude while $\chi 
(T)$ remains unaffected.  We refer to this phase as the 
low-temperature insulating (LI) phase.  The transition from the liquid 
to the crystal of bipolarons is a kind of Verwey 
transition~\cite{SchelenkerPMB1980,ChakravertyPMB1980} and the 
bipolaron-liquid formation in the HI phase is a kind of a SRO of 
charge carriers.  The Verwey transition~\cite{VerweyPhysica1941} was 
originally found for Fe$_{3}$O$_{4}$ at $T_{V} \simeq 120$ K. 
Analogous to Ti$_{4}$O$_{7}$, Fe$_{3}$O$_{4}$ has two possible valence 
states of Fe$^{2+}$ and Fe$^{3+}$ which are ordered below $T_{V}$ and 
are disordered above $T_{V}$, leading to the characteristic jump in 
the electrical resistivity.

In this Letter, we have studied the single-particle excitation spectra 
of the three phases of Ti$_{4}$O$_{7}$ by means of photoemission 
spectroscopy (PES) with high energy resolution.  To add to the 
variation of the electronic structure of Ti$_{4}$O$_{7}$ as a function 
of temperature across $T_{MI}$ and $T_{II}$, we focus on general 
aspects of the single-particle excitation spectra of dynamically 
disordered systems and then made comparison with the PES result of 
Fe$_{3}$O$_{4}$~\cite{ChainaniPRB1995}.

Single crystals of Ti$_{4}$O$_{7}$ were grown by the vapor transport 
method~\cite{SinceJCG1977}.  PES measurements were performed using an 
OMICRON hemi-spherical analyzer and a He lamp (He I: $h\nu=$21.2 eV).  
The energy calibration and the estimation of the instrumental 
resolution were done by measuring the Fermi edge of Au evaporated on 
the sample.  The energy resolution was set $\sim$50 meV. Samples were 
cleaved \textit{in situ}.  This always gave an irregular surface, an 
assembly of randomly orientated small facets.  As the analyzer had an 
acceptance angle of $\pm 8^{\circ}$, the obtained spectra can be 
regarded as angle-integrated PES spectra.  The measurement temperature 
was controlled within the accuracy of $\pm 0.2$ K. The base pressure 
in the spectrometer was less than $1 \times 10^{-10} $ Torr.  Below, 
we show results reproducible for several cleaves and obtained within a 
few hours after cleavage.

The position and the width of the O 2\textit{p} band (not shown) show 
almost no temperature dependence across the two phase transitions as 
in the previous photoemission study~\cite{AbbatePRB1995}.  In 
contrast, the PES spectra in the Ti 3\textit{d} band region show 
strong temperature dependence as shown in Fig.~\ref{Ti4O7narrow} (b), 
being consistent with the previous results for the M and LI phases 
with a lower energy resolution~\cite{AbbatePRB1995}.  Here, the sample 
was first cooled and then heated as indicated in the figure, so as to 
cross twice each transition temperature.  Two observations are 
remarked.  First, as superimposed in the bottom panel, there are three 
kinds of spectra, which reflect the electronic structure of the M, HI, 
and LI phases of Ti$_{4}$O$_{7}$ as discussed below.  Second, the 142 
K spectra show different spectral lineshapes between cooling and 
heating due to thermal hysteresis between the HI and LI phases, as has 
been observed in the electrical resistivity, 
thermopower~\cite{SchlenkerPDM1985}, and EPR 
measurements~\cite{LakkisPRB1976}.  In Fig.~\ref{Ti4O7narrow} (c), we 
have plotted the temperature dependence of the integrated PES 
intensity within 0.5 eV of $E_{F}$ normalized to the intensity 
integrated from $E_{F}$ to binding energy $E_{B} = 1$ eV. This 
hysteretic behavior was reproducible for several thermal cyclings 
across $T_{MI}$ and $T_{II}$.

\begin{figure} [ht] \center \epsfxsize=83mm \epsfbox{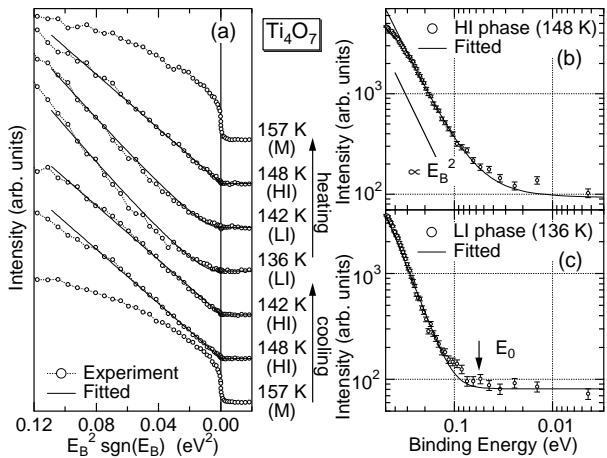} 
\vspace{1pt} \caption{(a) PES spectra of Ti$_{4}$O$_{7}$ plotted 
against $E_{B}^{2}\sgn(E_{B})$ with the fitted curves.  The 
experimental spectrum and the fitted curve plotted on the log-log 
scale for the HI phase (b) and the LI phase (c).}
\label{Ti4O7narrowFit}
\end{figure}

Let us discuss the spectra of each phase in more detail.  As shown in 
Fig.~\ref{Ti4O7narrow} (d), M-phase spectra show a weak but finite 
Fermi edge, reflecting its metallic nature.  The spectrum shows a 
broad maximum centered at $E_{B} =0.75$ eV, around which most of 
spectral weight is distributed.  A similar broad feature has been 
observed at $E_{B} \sim 1$--1.5 eV in the spectra of three-dimensional 
titanium and vanadium oxides with $3d^{1}$ configuration, LaTiO$_{3}$ 
and YTiO$_{3}$~\cite{FujimoriPRL92}, and has been interpreted as the 
incoherent part of the spectral function accompanying the coherent 
quasi-particle excitations around $E_{F}$.  The coherent part in 
Ti$_{4}$O$_{7}$ is hardly separable from the incoherent part because 
of their strong overlap, leading to the pseudo-gap-like behavior at 
$E_{F}$.  This implies that the M phase of Ti$_{4}$O$_{7}$ is a 
strongly correlated metal, where the motion of conduction electrons is 
largely incoherent, resulting in the weak coherent part.  In going 
from the M to the LI phase, the incoherent peak at $E_{B} = 0.75$ eV 
becomes sharper and a clear gap of order $\sim$0.1 eV is opened as 
seen in Fig.~\ref{Ti4O7narrow} (d).  The gap opening is attributed to 
the bipolaron ordering in this phase.  The HI-phase spectra take an 
intermediate lineshape between the LI and M phases: The 0.75 eV peak 
is somewhat broader than in the LI phase.  The spectra show neither a 
Fermi edge nor a real gap.  Indeed, the PES intensity vanishes only at 
$E_{B}=0.0$ eV as shown in Fig.~\ref{Ti4O7narrow} (d).  
Figure~\ref{Ti4O7narrowFit} (a) shows the same spectra plotted against 
$E_{B}^{2}\sgn(E_{B})$.  The Fermi edge in the M-phase spectra is now 
clearer while the LI spectra, which show a ``hard gap'', are concave 
around $E_{F}$.  Most remarkably, the HI-phase spectra form an almost 
straight line from $E_{B}^{2} = 0.00$ to $\sim 0.08$ eV$^{2}$ ($0.0 
\leq E_{B} \lesssim 0.3$ eV), meaning that the spectra show a 
power-law behavior with the exponent of $\sim 2$.

To be more quantitative, we have performed a lineshape analysis for 
the LI- and HI-phase spectra taking the experimental resolution into 
account.  The lineshape was assumed to be of the form $I(E_{B}, T) = 
I_{0}(E_{B}, T)+ I_{bg}(E_{B})$, where $I_{0}(E_{B}, T)$ is the 
intrinsic part expressed by $I_{0}(E_{B}, T) = a_{2}E_{B}^{2}f(E_{B}, 
T)$.  $f(E_{B}, T) = [\exp(-{E_{B}/k_{B}T})+1]^{-1}$ is the 
Fermi-Dirac distribution function at $T$ although the 
finite-temperature effect due to $f(E_{B}, T)$ is negligibly small 
because $I_{0}(0, T) = 0$.  $I_{bg}(E_{B})= b_{0}$ represents a 
constant background of the PES spectra.  For the LI phase, in order to 
represent the finite gap, we assumed $I_{0}(E_{B}, T) = 
a_{2}(E_{B}-E_{0})^{2}\heviside(E_{B}-E_{0})$, where $\heviside(x)$ is 
a step function and $E_{0}$ was allowed to take a finite positive 
value.  The instrumental resolution was included through the 
convolution of $I(E_{B}, T)$ with a Gaussian of FWHM $50$ meV. $a_{2}$ 
and $b_{0}$ were treated as adjustable parameters.  As shown in 
Fig.~\ref{Ti4O7narrowFit} (a), the result of the fitting is 
satisfactory, especially for the HI-phase spectra for $-0.1 \leq E_{B} 
\leq 0.3$ eV. One HI-phase spectrum is shown in 
Fig.~\ref{Ti4O7narrowFit} (b) on the log-log scale to emphasize the 
$E_{B}^{2}$ dependence and the excellent fit for small $E_{B}$.  This 
makes a clear contrast to the LI-phase spectrum shown in 
Fig.~\ref{Ti4O7narrowFit} (c), which shows a finite gap of $E_{0}$.

We propose that the above behavior of the HI phase can be explained as 
an SCG in the DOS, $N(E_{B}) = 
\frac{3}{\pi}{(\frac{\kappa}{e^{2}})}^{3}E_{B}^{2}$%
~\cite{EfrosJPC1975}, where $\kappa$ is the dielectric constant.  For 
Ti$_{4}$O$_{7}$, the magnitude of the SCG $\Delta_{C}= 
e^{3}(N_{0}/{\kappa^{3}})^{1/2}$, where $N_{0}$ is the noninteracting 
DOS at $E_{F}$, is estimated to be $\sim$0.2 eV if we take $N_{0} \sim 
0.01$ eV$^{-1}${\AA}$^{-3}$ from the free-electron-like DOS around 
$E_{F}$ and $\kappa \sim 10$.  As the estimated $\Delta_{C}$ and the 
observed soft gap have the same order of magnitude, we may conclude 
that the HI-phase spectra are consistent with the opening of an SCG. 
Here, it should be noted that, even if there exists substantial SRO, 
i.~e., bipolaron-liquid formation, in the HI phase of Ti$_{4}$O$_{7}$ 
the system is sufficiently disordered over long distance for an SCG to 
appear around $E_{F}$.  That is, the length scale of the SRO, which is 
equal to the Ti-Ti distance $\sim$3 {\AA}~\cite{MarezioPRL1972}, is 
sufficiently shorter than the critical distance $R_{C} \sim 
e^{2}/\kappa \Delta_{C}\sim 8$ {\AA} for the SCG to survive.  The SRO 
would be reflected on high-energy spectral features, e.~g., the 
sharpening of the $E_{B} = 0.75$ eV peak in the HI phase.  
Experimentally, an SCG was observed in tunneling spectroscopy 
measurements of the doped semiconductor Si:B~\cite{MasseyPRL1996}, 
whose gap was as small as 0.75 meV due to low $N_{0}$.  As for PES 
measurements, besides the aforementioned 
Na$_{x}$Ta$_{y}$W$_{1-y}$O$_{3}$, possible existence of SCG in 
Ti$_{4}$O$_{7}$ and Fe$_{3}$O$_{4}$ was 
suggested~\cite{DaviesPRL1986}, while no quantitative analysis on the 
experimental spectra has been performed so far.  In this sense, the 
present result $I \propto E_{B}^{2}$ for the HI phase of 
Ti$_{4}$O$_{7}$ is the first clear indication of an SCG using PES. 
Since our measurements were performed on cleaved surfaces, there are 
no extrinsic disorder effects induced by scraping.

\begin{figure} [ht] \center \epsfxsize=80mm \epsfbox{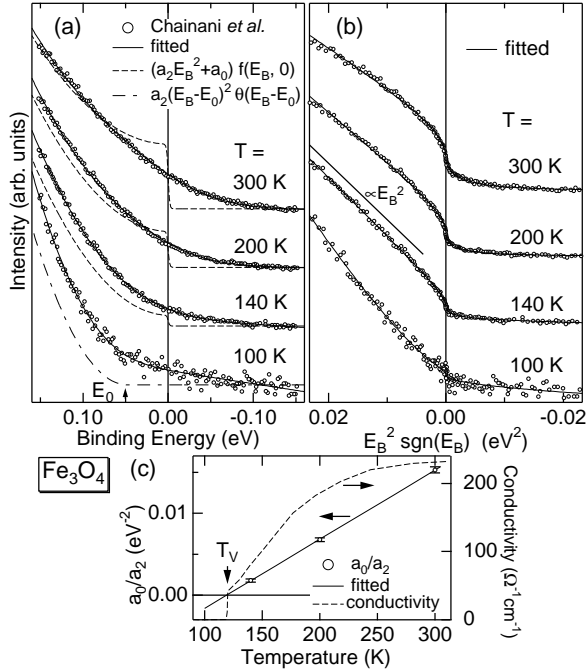} 
\vspace{1pt} \caption{(a) PES spectra of Fe$_{3}$O$_{4}$ ($h\nu = 
21.2$ eV) taken from Ref.~\protect\cite{ChainaniPRB1995}.  Solid, 
dashed, and dotted-dashed curves represent the fitted curves, 
$(a_{2}E_{B}^{2}+a_{0})f(E_{B},0)$, and 
$a_{2}(E_{B}-E_{0})^{2}\heviside(E_{B}-E_{0})$, respectively.  (b) The 
same spectra plotted against $E_{B}^{2}\sgn(E_{B})$.  (c) Ratio 
$a_{0}/a_{2}$, which tends to vanish at $T \sim T_{V}$ (see text).  
The electrical conductivity of Fe$_{3}$O$_{4}$ is also shown by a 
dashed line~\protect\cite{ParkPRB1998}.  }
\label{Fe3O4fig}
\end{figure}

It is worth comparing Ti$_{4}$O$_{7}$ in the HI phase with 
Fe$_{3}$O$_{4}$ above $T_{V} \simeq 120$ K because of the analogy 
between the two materials pointed out 
repeatedly~\cite{ChakravertyPMB1980}.  For this purpose, we reanalyzed 
the PES spectra of Fe$_{3}$O$_{4}$ reported by Chainani \textit{et 
al.}~\cite{ChainaniPRB1995} in the context of an SCG. 
Figure~\ref{Fe3O4fig} (a) shows the PES spectra of Fe$_{3}$O$_{4}$ 
measured with the resolution of $\sim$70 meV. They claimed that the 
intensity at $E_{F}$ in the metallic phase evolves as the temperature 
increases.  In Fig.~\ref{Fe3O4fig} (b), we have replotted the spectra 
against $E_{B}^{2}\sgn(E_{B})$, which makes their argument clearer.  
More interestingly, we find that the 140 K ($>T_{V}$) spectrum almost 
falls on a straight line at $E_{B} \gtrsim 0.07$ eV. In contrast, the 
100 K spectrum is slightly concave, signaling the opening of a hard 
gap.  Fe$_{3}$O$_{4}$ is, however, different from Ti$_{4}$O$_{7}$ in 
that the 140 K spectrum, which would correspond to the HI-phase 
spectra of Ti$_{4}$O$_{7}$, does not vanish at $E_{F}$ but shows a 
finite Fermi edge.  We have modeled the spectra $I(E_{B}, T)$ above 
$T_{V}$ as $I(E_{B}, T) = I_{0}(E_{B}, T)+ I_{bg}(E_{B})$, where 
$I_{0}(E_{B}, T) = (a_{2}E_{B}^{2}+a_{0}) f(E_{B}, T)$ ($a_{0}$: 
finite DOS at $E_{F}$).  The 100 K spectrum was assumed to be 
$I_{0}(E_{B}, T) = a_{2}(E_{B}-E_{0})^{2}\heviside(E_{B}-E_{0})$ as in 
the LI phase of Ti$_{4}$O$_{7}$.  $I_{bg}(E_{B}) = b_{1}E_{B}+b_{0}$ 
represents the linear background of the PES 
spectra~\cite{CommnetOnBG}.  The instrumental resolution was also 
included.

The result of the fitting was satisfying between $E_{B} = -0.15$ eV 
and 0.15 eV as shown by solid curves in Figs.~\ref{Fe3O4fig} (a) and 
(b).  The dashed curves in Fig.~\ref{Fe3O4fig} (a) for $T \geq 140$ K 
represent the intrinsic DOS cut-off at $E_{F}$ 
$(a_{2}E_{B}^{2}+a_{0})f(E_{B}, 0)$.  The most remarkable point is 
that the finite intensity at $E_{F}$ evolves systematically with 
increasing temperature.  Figure~\ref{Fe3O4fig} (c) shows the ratio 
$a_{0}/a_{2}$ between the intensity at $E_{F}$, which contributes to 
the ``metallic'' conductivity, and the magnitude of the $E_{B}^{2}$ 
term, which represents the dynamical disorder.  The electrical dc 
conductivity data of Fe$_{3}$O$_{4}$~\cite{ParkPRB1998} is 
superimposed in the figure.  The $a_{0}/a_{2}$ values at 140, 200, and 
300 K form a straight line which extrapolates to 0 at 119$\pm 5$~K 
$\sim T_{V}$.  This observation, namely, $a_{0}/a_{2} \propto T-T_{V}$ 
strongly indicates that $I_{0} \propto E_{B}^{2}$ just above $T_{V}$.  
As the temperature goes up above $T_{V}$, the finite intensity at 
$E_{F}$ grows up in proportion to $T-T_{V}$.  Park \textit{et 
al.}~\cite{ParkPRB1997} reported that the spectra of Fe$_{3}$O$_{4}$ 
at 130 K show no Fermi edge, which is naturally understood as due to 
the small $a_{0}/a_{2}$.  We propose that an SCG exists in 
Fe$_{3}$O$_{4}$ just above $T_{V}$ and continuously evolves into a 
pseudogap well above $T_{V}$, reflecting the crossover from the 
semiconducting ($d\rho/dT < 0$) behavior just above $T_{V}$ to the 
metallic ($d\rho/dT > 0$) one well above $T_{V}$.  Just above $T_{V}$, 
charges are disordered with SRO, namely, the system is in a Wigner 
glass state~\cite{MottMIT} as is the HI phase of Ti$_{4}$O$_{7}$.  
Here, it should be noted that the M-phase spectra of Ti$_{4}$O$_{7}$ 
near $E_{F}$ ($E_{B} < 0.2$ eV) may be fitted to 
$(a_{2}E_{B}^{2}+a_{0})f(E_{B}, T)$ like the spectra of 
Fe$_{3}$O$_{4}$ above $T_{V}$.  However, $a_{0}/a_{2} = 2$-4$\times 
10^{-2}$ eV$^{-2}$ is much larger than that in Fe$_{3}$O$_{4}$ at 300 
K, reflecting the strongly first-order transition between the HI and M 
phases in Ti$_{4}$O$_{7}$.

In conclusion, we have performed a PES study of Ti$_{4}$O$_{7}$ 
covering its LI, HI, and M phases.  The spectra of the Ti 3\textit{d} 
band show peculiar temperature-dependent spectra characteristic of the 
three phases, among which the HI-phase spectra are gapless and can be 
fitted to $E_{B}^{2}$ near $E_{F}$.  We interpret this as an SCG 
resulting from disordered bipolarons.  By reanalyzing the PES spectra 
of Fe$_{3}$O$_{4}$, an SCG was also found just above $T_{V}$, 
indicating the significant role of disorder and long-range Coulomb 
interaction.  With increasing temperature, the SCG is found to 
continuously evolve into a pseudogap as the metallicity is gradually 
established, whereas the first-order HI-to-M transition in 
Ti$_{4}$O$_{7}$ precludes the observation of the corresponding 
continuous change.  Presumably, similar SCG behavior may be observed 
in other systems such as 
$\beta$-Na$_{x}$V$_{2}$O$_{5}$~\cite{SchlenkerPDM1985}.  The very 
recent scanning tunneling spectra of hole-doped manganites above 
$T_{C}$~\cite{BiswasPRB1999} may be interpreted in the same way as 
Fe$_{3}$O$_{4}$ above $T_{V}$.

We would like to thank D. Khomskii, M. Abbate, and T. Mizokawa for 
informative discussions.  We appreciate Y. Aiura, K. Tobe and Y. 
Ishikawa for technical support.  This work was supported by the New 
Energy and Industrial Technology Development Organization (NEDO) and 
by a Special Coordination Fund from the Science and Technology Agency 
of Japan.


\begin{references}

\bibitem[*]{presentaddress}Present Address: Institute for Solid State 
Physics, University of Tokyo, Tokyo 106-8666, Japan.

\bibitem{MottMIT} N. F. Mott, \textit{Metal-Insulator Transitions}
(Taylor \& Francis, London, 1990).

\bibitem{AndersonPRL1975} P. W. Anderson, Phys.  Rev.  Lett.  {\bf 
34,} 953 (1975).

\bibitem{EfrosJPC1975}A. L. Efros and B. I. Shklovskii, J. Phys.  C 
{\bf 8,} L49 (1975).

\bibitem{MasseyPRL1996} J. G. Massey and M. Lee, Phys.  Rev.  Lett.  
{\bf 77,} 3399 (1996); {\bf 75,} 4266 (1995).

\bibitem{DaviesPRL1986} J. H. Davies and J. R. Franz, Phys.  Rev.  
Lett.  {\bf 57,} 475 (1986).

\bibitem{HollingerPRB1985} G. Hollinger \textit{et al.}, Phys.  Rev.  
B {\bf 32,} 1987 (1985); M. D. Hill and R. G. Egdell, J. Phys.  C {\bf 
16,} 6205 (1983).

\bibitem{SchlenkerPDM1985} C. Schlenker, \textit{Physics of Disordered 
Materials}, eds.\ by D. Alder, H. Fritzsche, and S. Ovshinsky (Plenum, 
New York, 1985) p.~369.

\bibitem{MarezioPRL1972} M. Marezio \textit{et al.}, Phys.  Rev.  Lett.  
{\bf 28,} 1390 (1972).

\bibitem{SchlenkerPRL1974} C. Schlenker \textit{et al.}, Phys.  Rev.  
Lett.  {\bf 32,} 1318 (1974).

\bibitem{LakkisPRB1976} S. Lakkis \textit{et al.}, Phys.  Rev.  B {\bf 
14,} 1429 (1976).

\bibitem{SchelenkerPMB1980} C. Schlenker and M. Marezio, Philos.  
Mag.  B {\bf 42,} 453 (1980).

\bibitem{ChakravertyPMB1980} B. K. Chakraverty, Philos.  Mag.  B {\bf 
42,} 473 (1980).

\bibitem{VerweyPhysica1941} E. J. W. Verwey and P. W. Haaymann, 
Physica {\bf 8,} 979 (1941).

\bibitem{ChainaniPRB1995} A. Chainani \textit{et al.}, Phys.  Rev.  B 
{\bf 51,} 17976 (1995).

\bibitem{SinceJCG1977} J. J. Since, S. Ahmed, and J. Mercier, J. 
Crystal Growth {\bf 40,} 301 (1977).

\bibitem{AbbatePRB1995} M. Abbate \textit{et al.}, Phys.  Rev.  B {\bf 
51,} 10150 (1995).

\bibitem{FujimoriPRL92} A. Fujimori \textit{et al.}, Phys.  Rev.  
Lett.  {\bf 69,} 1796 (1992); I. H. Inoue \textit{et al.}, 
\textit{ibid.} {\bf 74,} 2539 (1995).

\bibitem{CommnetOnBG} Because of the very low intrinsic photoemission 
signals around $E_{F}$ for Fe$_{3}$O$_{4}$, the linear background due 
to satellites ($h\nu = 23$--24 eV) of the He I light had to be taken 
into account.  Unlike Fe$_{3}$O$_{4}$, the satellites are negligibly 
weak in the spectra of Ti$_{4}$O$_{7}$ around $E_{F}$.

\bibitem{ParkPRB1998} S.-K. Park, T. Ishikawa, and Y. Tokura, Phys.  
Rev.  B {\bf 58}, 3717 (1998).

\bibitem{ParkPRB1997} J.-H. Park \textit{et al.}, Phys.  Rev.  B {\bf 
55,} 12813 (1997).

\bibitem{BiswasPRB1999} A. Biswas \textit{et al.}, Phys.  Rev.  B {\bf 
59}, 5368 (1999).
\end{references}
\end{document}